\documentclass[5p,times]{elsarticle}
\usepackage{epsfig,amsmath,amssymb}

\newcommand*{\TITLE}[1]{\section{#1}}
\newcommand*{\SUBTITLE}[1]{\subsection{#1}}
\newcommand*{\be}[0]{\begin{equation}}
\newcommand*{\ee}[0]{\end{equation}}
\newcommand*{\beu}[0]{\begin{equation}}
\newcommand*{\eeu}[0]{\end{equation}}
\newcommand*{\ba}[0]{\begin{array}}
\newcommand*{\ea}[0]{\end{array}}
\newcommand*{\bfig}[0]{\begin{figure}[ht]}
\newcommand*{\efig}[0]{\end{figure}}
\newcommand*{\bfigwide}[0]{\begin{figure*}}
\newcommand*{\efigwide}[0]{\end{figure*}}
\newcommand*{\fig}[1]{Figure~\protect\ref{fig:#1}}
\newcommand*{\tab}[1]{Table~\protect\ref{table:#1}}
\newcommand*{\eqn}[1]{Equation~\protect\ref{eqn:#1}}
\newcommand*{\scrE}[0]{\mathcal{E}}
\newcommand*{\dtCRSS}[0]{\Delta\tau_{\text{CRSS}(0001)}}
\newcommand*{\size}[0]{\varepsilon_b}
\newcommand*{\chem}[0]{\varepsilon_\text{SFE}}
\newcommand*{\conc}[0]{c_\text{s}}	%{c}

\newlength{\wholefigwidth}
\newlength{\smallfigwidth}
\newlength{\halfsmallfigwidth}
\journal{Acta Materialia}
\begin{document}
\begin{frontmatter}

\title{First-principles data for solid-solution strengthening of magnesium:
From geometry and chemistry to properties}

\author[UIUCPhysics]{Joseph A. Yasi}
\author[GM]{Louis G. Hector, Jr.}
\author[UIUCMatSE]{Dallas R. Trinkle}
\ead{dtrinkle@illinois.edu}

\address[UIUCPhysics]{Department of Physics, University of Illinois at
Urbana-Champaign, Urbana, IL 61801}
\address[GM]{General Motors R\&D Center, 30500 Mound Road, Warren, MI 48090}
\address[UIUCMatSE]{Department of Materials Science and Engineering, University of
Illinois at Urbana-Champaign, Urbana, IL 61801}

\begin{abstract}
Solid-solution strengthening results from solutes impeding the glide of
dislocations.  Existing theories of strength rely on solute-dislocation
interactions, but do not consider dislocation core structures, which need
an accurate treatment of chemical bonding.  Here, we focus on strengthening
of Mg, the lightest of all structural metals and a promising replacement
for heavier steel and aluminum alloys.  Elasticity theory, which is
commonly used to predict the requisite solute-dislocation interaction
energetics, is replaced with quantum-mechanical first-principles
calculations to construct a predictive mesoscale model for solute
strengthening of Mg.  Results for 29 different solutes are displayed in a
``strengthening design map'' as a function of solute misfits that quantify
volumetric strain and slip effects.  Our strengthening model is validated
with available experimental data for several solutes, including Al and Zn,
the two most common solutes in Mg.  These new results highlight the ability
of quantum-mechanical first-principles calculations to predict complex
material properties such as strength.
\end{abstract}


\end{frontmatter}

%%%%%%%%%%%%%%%%%%%%%%%%%%%%%%%%%%%%%%%%%%%%%%%%%%%%%%%%%%%%%%%%%%%%%%%%
\TITLE{Introduction}

Inherent limitations of strength and formability, which are related to
microscopic-scale deformation mechanisms, are significant obstacles to
widespread adoption of wrought magnesium in transportation industries.
Magnesium's poor workability at room temperature comes from the anisotropic
response of its hexagonal closed-packed (HCP) crystal structure, and the
strength of conventional Mg alloys is lower than that of most aluminum
alloys without added processing steps (e.g. grain refinement).  While the
polycrystalline ductility of face- and body-centered cubic metals results
from multiple symmetry-related slip systems, the basal and prismatic planes
in Mg are perpendicular in its HCP lattice, unrelated by symmetry, and must
both be active to achieve appreciable ductility.  The room temperature
stress required to plastically deform Mg along its (easy) basal slip plane
is two orders-of-magnitude lower that the (hard) prismatic plane. The five
independent slip systems required by the von~Mises
criterion\cite{Taylor1938} for sufficient ductility are simultaneously
activated only at temperatures near $300^\circ\text{C}$.  Insights required
to overcome these design challenges can come from new predictive
capabilities.  To that end, we develop a new accurate first-principles
strengthening model that predicts Mg basal strengthening as a function of
substitutional solute chemistry via computation of the fundamental
solute-dislocation interaction.

We compute the interaction energy of solutes with screw and edge basal
dislocations---finding surprisingly similar magnitude interactions of
solutes with both dislocations types---and use this information to build a
first-principles ``design map'' for the strengthening of solutes in Mg in a
computationally efficient manner.  Using chemically accurate predictions of
the atomic-scale geometry, we resolve the volumetric expansion and
compression and the local slip in both the far-field and into the
dislocation cores, and determine the solute-dislocation interaction energy
for all substitution sites.  This combines accurate dislocation core
geometries at the atomic scale---available from state-of-the-art
first-principles
calculations\cite{vasp1993,vasp1996,perdew1992,vanderbilt1990,kresse1994}
with flexible boundary condition methods\cite{rao1998,Trinkle2008}, proven
successful for Mo\cite{Trinkle2005} and Al\cite{Woodward2008}---and
computation of solute ``misfits.''  The misfit of a solute quantifies
relative changes in the lattice from dilute substitution of a solute, and
consists of two important components: a ``size'' misfit for the change in
the local volume, and a ``chemical'' misfit for the change in energy to
slip the crystal in the basal plane.  The misfits provide the basis for
approximating the solute-dislocation interaction energy; we test this
approximation against direct substitution of Al in Mg dislocations before
applying it for other solutes.  The solute/screw dislocation interaction
energy, normally assumed to be negligible in the far-field from elasticity
treatments\cite{Neuhauser1993,HullBacon2001} but which has recently been
quantified with atomic-scale studies\cite{Trinkle2005,Olmsted2005}, is
found to be nearly \textit{identical} in magnitude to the edge dislocation
interaction energy due to the screw core geometry and contributes to the
prediction of strength.  Finally, we use our geometrically-informed
calculation of solute-dislocation interactions to predict the
dilute-concentration strengthening effect of 29 different solutes and
create a simple map of strengthening potencies that suggests new Mg alloy
designs.  Our entirely first-principles approach is validated with
available experimental strengthening data.

%%%%%%%%%%%%%%%%%%%%%%%%%%%%%%%%%%%%%%%%%%%%%%%%%%%%%%%%%%%%%%%%%%%%%%%%
\TITLE{Computational methods}

\bfig
\centering
\includegraphics[width=\smallfigwidth]{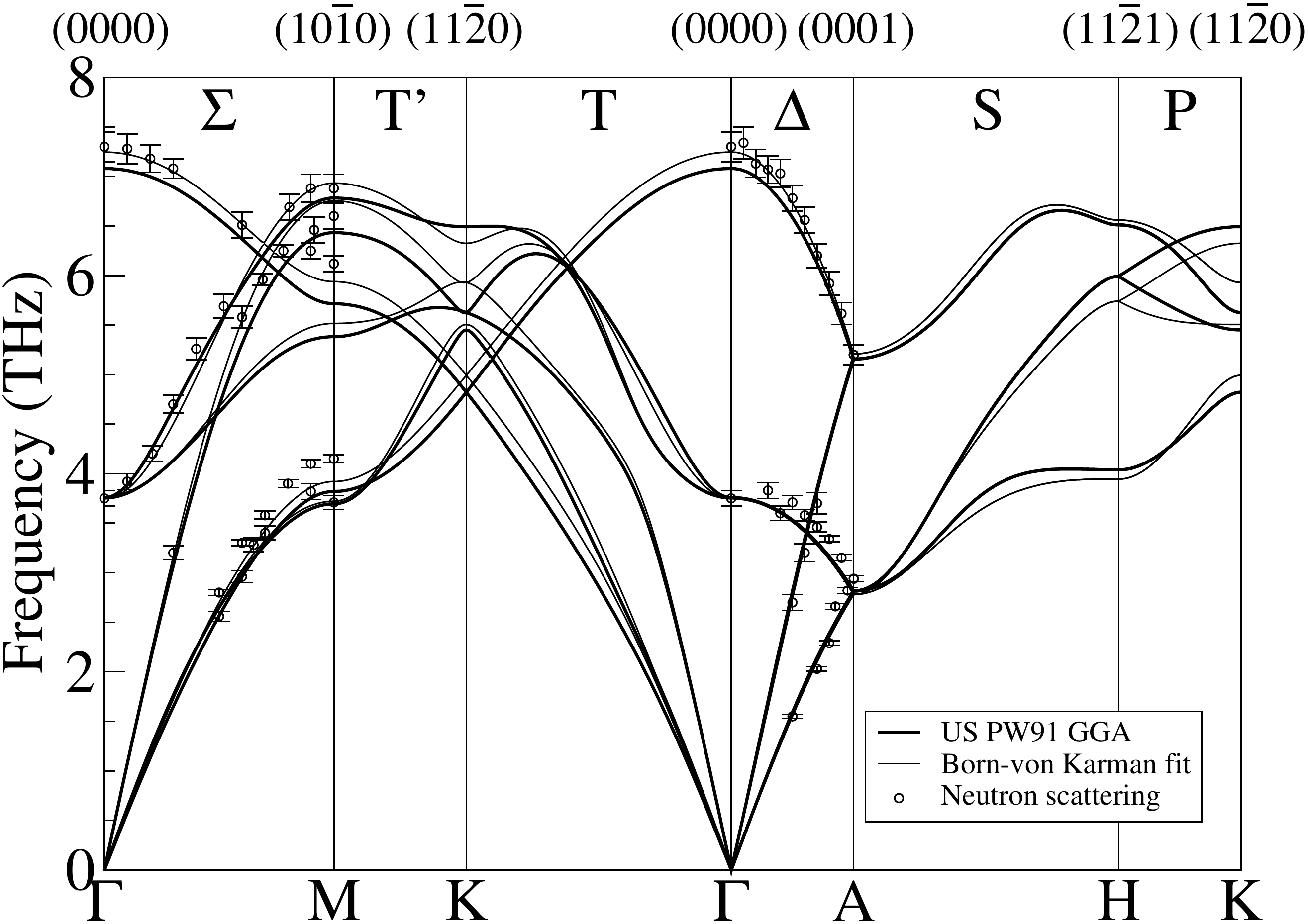}
\includegraphics[width=\smallfigwidth]{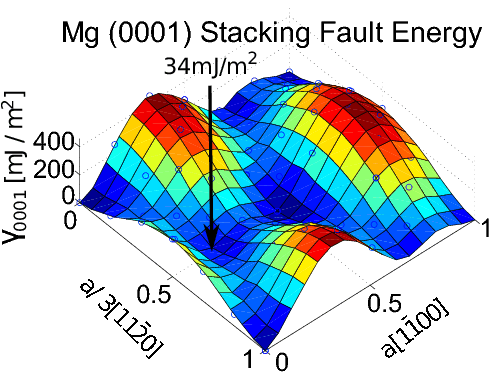}

\caption{First-principles phonon spectra (top) and $(0001)$ generalized
stacking-fault energy (bottom) for Mg.  These describe response of the
lattice to small displacements, elastic deformation, and slip in the basal
plane.  Density-functional theory is able to accurately reproduce the
vibrational spectra from experimental measurements: Neutron scattering data
from \protect\cite{Iyengar1965}, and a Born-von~Karman fit to the data from
\protect\cite{Pynn1972}.  The equilibrium lattice constants from
the ultrasoft pseudopotential are $a=3.19\text{\AA}$ and
$c=5.18\text{\AA}$; the elastic constants, which determine the slopes of
the spectra for long wavelengths are $C_{11}=60\text{GPa}$,
$C_{33}=61\text{GPa}$, $C_{12}=21\text{GPa}$, $C_{13}=20\text{GPa}$, and
$C_{44}=18\text{GPa}$.  For the generalized-stacking fault surface, a
single plane of Mg is displaced in the basal plane by a linear combination
of $\frac{a}{3}[11\bar20]$ and $a[1\bar100]$; the defected geometry is
allowed to relax in the $[0001]$ directions, and the energy per area for
the defect is the generalized stacking-fault energy.  The highlighted point
of $\frac{a}{3}[10\bar10]$ is a metastable configuration known as the
intrinsic I2 stacking fault.  The $\gamma_\text{I2}$ energy agrees well
with other density-functional theory calculations of the
same\protect\cite{Chetty1997}.  The I2 stacking fault geometry is the basis
for computing chemical misfits for solutes.}
\label{fig:phononGSFE}
\efig

Efficient computation of interactions between solutes and dislocations
requires simulation cells and flexible boundary condition approaches that
isolate individual defects.  Calculations are performed with
VASP\cite{vasp1993,vasp1996}, a plane-wave density-functional theory code.
Magnesium and all solutes are treated with Vanderbilt ultrasoft
pseudopotentials\cite{vanderbilt1990,kresse1994}, and the Perdew-Wang 91
GGA exchange-correlation potential\cite{perdew1992}.  The ultrasoft Mg
pseudopotential ([Ne]$3s^2$) accurately reproduces experimental
lattice\cite{Errandonea2003} (less than 0.9\%\ error) and
elastic\cite{Schmunk1959} (less than 5\%\ error) constants, and phonon
frequencies\cite{Iyengar1965,Pynn1972} (less than 3\%\ error) of bulk Mg;
c.f.~\fig{phononGSFE}.  The stacking-fault surface in \fig{phononGSFE},
while not available experimentally, compares well with other
density-functional theory calculations\cite{Chetty1997}.  We chose a
planewave cutoff of 138eV, $k$-point meshes (see below for specific values
tied to each geometry) and Methfessel-Paxton smearing of 0.5eV to give an
energy accuracy of 5meV for bulk Mg.  For calculations involving solutes,
the cutoff energy was increased as necessary to accommodate harder
pseudopotentials for solutes; c.f.~\tab{solute} for all cutoff energies
used for the corresponding pseudopotentials (in general, we selected a
cutoff energy of 1.3 times the suggested cutoff for the potential).  The
geometries for misfits, and dislocation calculations and coupling with
lattice Green function flexible boundary condition methods are given below.

\SUBTITLE{Size misfit}
For the size misfit, we substituted single solutes into a $2\times2\times2$
Mg supercell ($k$-point mesh of $16\times16\times10$) at five different
volumes based on the equilibrium Mg volume $V_0$: $1.16V_0$, $1.05V_0$,
$1.00V_0$, $0.95V_0$, $0.86V_0$.  The atomic positions in each supercell
were relaxed until all forces were less than 5meV/\AA.  The size misfit is
the logarithmic derivative of the Burgers vector $b=\frac{a}{3}[2\bar110]$
with solute concentration $\conc$ in the dilute limit; hence,
\beu
\size = \left.\frac{d \ln b}{d\conc}\right|_{\conc=0}
= \frac{1}{b}\left.\frac{db}{d\conc}\right|_{\conc=0}.
\eeu
The change in energy for a solute with volume is given by the differences
in supercell energies $E_\text{solute supercell}(e_V) - E_\text{solute
supercell}(0) - 15 E_\text{bulk Mg}(e_V) + 15 E_\text{bulk Mg}(0)$ and
varies with the volumetric strain $e_V = V/V_0 - 1$ for each supercell.
The raw data is fit to a quadratic in strain $E'_\text{solute} e_V +
E''_\text{solute} e_V^2$, and the slope $E'$ is used to determine the size
misfit
\beu
\size = -\frac{E'}{3BV_0},
\eeu
where $B$ is the bulk modulus.  We use the slope $E'_\text{solute}$ to
compute the change in binding energy for a solute due to local volumetric
strain.  Comparison with calculations using a larger $3\times3\times3$
supercell affected the size misfit by $\lesssim 10\%$.

\SUBTITLE{Chemical misfit}
For the chemical misfit, we substituted single solutes into a
$2\times2\times9$ Mg supercell ($k$-point mesh of $16\times16\times1$)
corresponding to 18 $(0002)$ planes.  The supercell vector along the
$c[0001]$ direction has an extra $\frac{a}{3}[01\bar10]$ component, so that
the supercell represents a periodic repetition of stable stacking faults
separated by a distance of $9c$.  The solute is substituted into a site in
the stacking fault, and the atomic positions were relaxed until forces were
less than 5meV/\AA.  Because the I2 intrinsic stacking-fault configuration
is (meta)stable (c.f.~\fig{phononGSFE}), we do not impose any constraints
on the relaxation.  The chemical misfit is the logarithmic derivative of
the I2 intrinsic stacking-fault energy $\gamma_\text{I2}$ with solute
concentration $\conc$ in the dilute limit; hence,
\beu
\chem = \left.\frac{d\ln \gamma_\text{I2}}{d\conc}\right|_{\conc=0}
= \frac{1}{\gamma_\text{I2}} \left.
\frac{d\gamma_\text{I2}}{d\conc}\right|_{\conc=0}.
\eeu
The chemical misfit is calculated from our supercell energies as
\beu
\chem = 
\frac{E_\text{displaced}(\text{solute}) -
E_\text{undisplaced}(\text{solute}) -
2\sqrt{3}a^2\gamma_\text{I2}}{\gamma_\text{I2}\sqrt{3}a^2/2},
\eeu
where the ``displaced'' and ``undisplaced'' geometries correspond to the
layered structure with and without an I2 intrinsic stacking fault, and
$\sqrt{3}a^2/2$ is the basal plane area.

\SUBTITLE{Dislocation geometries}

\bfig
\centering
\includegraphics[width=\smallfigwidth]{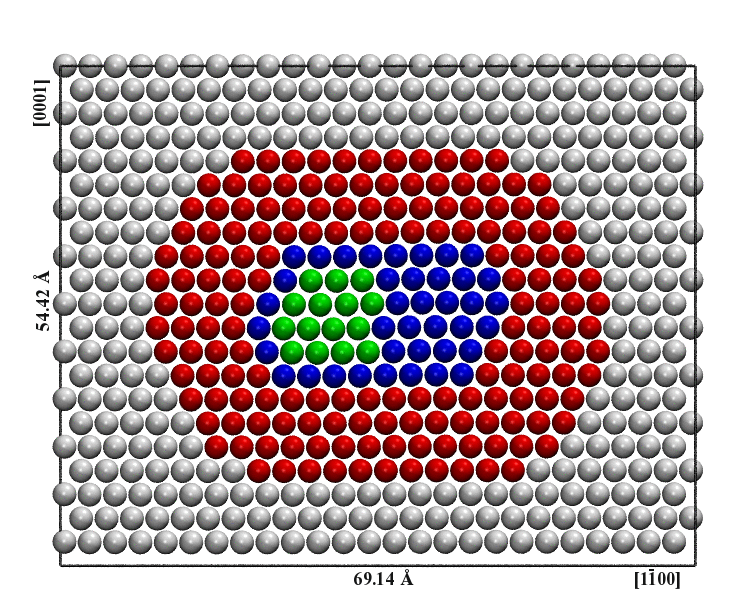}
\includegraphics[width=\smallfigwidth]{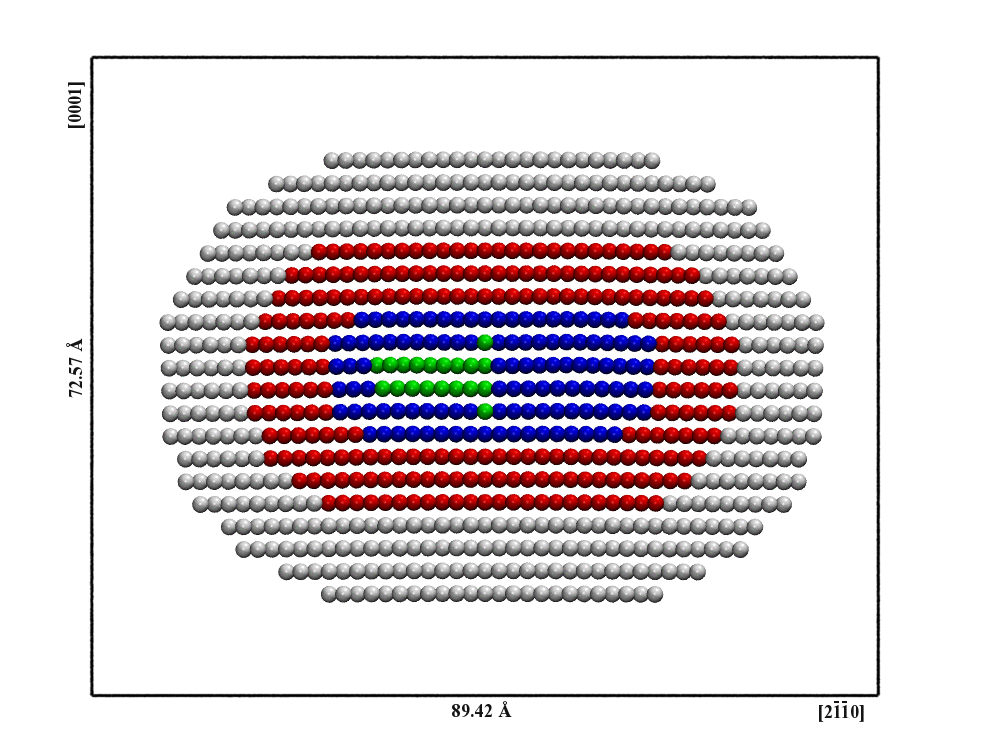}

\caption{Full Mg basal screw (top) and edge (bottom) dislocation core
equilibrium geometries, separated into regions I (blue), II (red), and III
(gray) and identified solute sites (green).  The periodic simulation boxes
are $69.14 \times 54.42 \times 3.19$ \AA$^3$ for screw and $84.92 \times
72.57 \times 5.53$ \AA$^3$ for edge.  The initial screw geometry has
displacements coming out of the page, while the edge displacements are in
the plane of the page; relaxation produces additional displacements in the
core of the partial dislocations.}
\label{fig:geom}
\efig

Flexible boundary condition methods\cite{sinclair1978,rao1998,Trinkle2008}
relax the pure Mg basal dislocation geometries.  The starting geometry
comes from the anisotropic elasticity solution for a dislocation using the
elastic and lattice constants from our first principles
calculations \cite{HirthLothe,Bacon1979}.  The elasticity solution
determines the displacement field as a continuum function; we start with a
bulk HCP lattice periodically repeated along the dislocation line
direction: $\frac{a}{3}[2\bar1\bar10]$ for the screw ($k$-point mesh of
$1\times1\times16$) and $a[01\bar10]$ for the edge ($k$-point mesh of
$1\times1\times12$).  The displacement field is centered between two
$(0002)$ atomic planes, and applied to every atom.  A finite sized
computational cell is produced by simulating only atoms within 10 lattice
planes from the estimated positions of the partial cores; c.f. \fig{geom}.
The distance from the core is determined by summing the distance to return
to small relative displacements (the size of region I), the distance for
the lattice Green function to match the elastic Green
function\cite{Trinkle2008} (the size of region II), and the distance from
an unrelaxed free surface to produce zero forces (the size of region III).
Together, this ensures that region II both does not have spurious forces
due to the vacuum outside and is sufficiently separated from the defect
core to follow the bulk harmonic response.  In the case of the screw
dislocation, the lack of a long-range volumetric strain field allows the
use of a periodic simulation cell with ``domain boundaries'' at the cell
boundaries.  For the edge dislocation, vacuum is needed to isolate periodic
images of the dislocations.

\SUBTITLE{Flexible boundary conditions}
Lattice Green function-based flexible boundary conditions isolate the
dislocation at the center of the simulation cell from the domain boundaries
or vacuum while displacing surrounding atoms as if the dislocation were
embedded in infinite bulk responding
harmonically\cite{sinclair1978,rao1998}.  The method relaxes each atom based
on its location within one of three different regions determined by the
distance from the dislocation core.  Overall, the screw dislocation
consists of 525 atoms and the edge dislocation consists 806 atoms.  Region
I atoms near the core (54 for the screw, 130 for the edge) start the
simulation with non-zero forces, and are relaxed using a conjugate-gradient
method.  As the relaxation commences, the neighboring atoms in region II
(164 for the screw, 247 for the edge) start with zero forces but the forces
increase as atoms in region I are displaced.  The region II atoms can be
treated as if they were coupled to infinite harmonic bulk; hence, their
forces can be relaxed by applying a displacement to each atom $\vec R$
given by the lattice Green function (LGF): $\vec u(\vec R) = \sum_{\vec R'}
{\bf G}(\vec R-\vec R') \vec f(\vec R')$, where $\vec R'$ only varies over
atoms in region II, $\vec f$ are their forces, and ${\bf G}(\vec R-\vec
R')$ is the (tensor) LGF.  This displaces all atoms---including region I
atoms, and the outer region III atoms (307 for the screw, and 429 for the
edge).  The atoms in the outer region have non-zero forces due to the
domain boundary or vacuum; but the lattice Green function displaces them as
if they were part of an infinite bulk lattice.  The relaxation cycle
continues between region I (conjugate gradient) and II (lattice Green
function) until the forces are less than 5meV/\AA\ in both regions.  The
final result of the relaxation is the stress-free dislocation core
equilibrium geometry.

\SUBTITLE{Direct solute-dislocation interaction energy calculation}
We compute the interaction of Al with Mg dislocations by direct
substitution of Al for Mg at different sites in the dislocation cores;
c.f. \fig{geom}.  For each substitution, the entire region I was relaxed
until the forces were less than 5meV/\AA.  This defines the relative
energies for Al in each site (18 for the screw, 30 for the edge); to define
the energy zero for Al---the reference of Al substituted into bulk Mg with
no strain field---we reference the average energy of Al above and below the
stacking fault region.  The periodic repetition of the solute along the
dislocation line introduces a finite-size error in the calculated
solute-dislocation interaction.  For the screw dislocation geometry, there
is one Al atom every 3.19\AA\ along the dislocation line, and for the edge
dislocation, every 5.53\AA.  Using a screw dislocation geometry with double
the periodicity (6.38\AA), the calculated Al solute/screw dislocation
interaction energy was 8.4meV higher than that extracted from the original
geometry.  We expect this to be an upper limit on the finite-size error as
it is (a) for the shortest Al-Al repeat distance, and (b) for the largest
change in local geometry.

\SUBTITLE{Misfit approximation of solute-dislocation interaction energy}
To compute the interaction of any solute with the Mg dislocations, we
analyzed each dislocation geometry in terms of local volumetric strain and
slip.  The local volumetric strain at each atomic site in the final relaxed
dislocation geometry is defined from the nearest-neighbor positions as
\beu
e_V = \left[\frac{
\det\left\{\sum_{\vec x'} x'_i x'_j\right\}}{
\det\left\{\sum_{\vec x} x_i x_j\right\}}
\right]^{1/2} - 1
\eeu
where $\vec x'$ are the vectors to nearest neighbors for a site, and $\vec
x$ are the corresponding nearest-neighbor vectors in the HCP
lattice\cite{Hartley2005}.  The slip interaction energy ($E_\text{slip}$)
is calculated at each atomic site as
\beu
E_\text{slip} =
\frac{\sqrt{3}a^2/2}{6}\sum_{\vec d}\gamma_{(0001)}\left(\vec d\right),
\eeu
where $\vec{d}$ are the vectors to the nearest neighbors in adjacent basal
planes, $\gamma_{(0001)}$ is the generalized basal stacking-fault energy
for displacement $\vec d$, and the factor of $\frac{1}{6}$ is from
assigning half the bond energy for the 3 out-of-plane neighbors.  The
interaction energy for a solute is then a sum of two contributions: the
size interaction (given by the change in solute energy at the site strain)
and slip interaction (given by the chemical misfit multiplied by the slip
energy of the site)
\be
\begin{split}
E_\text{binding} &= E'_\text{solute}e_V +  E_\text{slip}\cdot\chem\\
&= -3BV_0\cdot e_V\cdot\size + E_\text{slip}\cdot\chem.
\end{split}
\label{eqn:misfitbinding}
\ee
This misfit approximation requires significantly less computing resources
to determine compared with direct calculations.

%%%%%%%%%%%%%%%%%%%%%%%%%%%%%%%%%%%%%%%%%%%%%%%%%%%%%%%%%%%%%%%%%%%%%%%%
\TITLE{Dislocation / solute interactions}

\bfig
\centering
\includegraphics[width=\smallfigwidth]{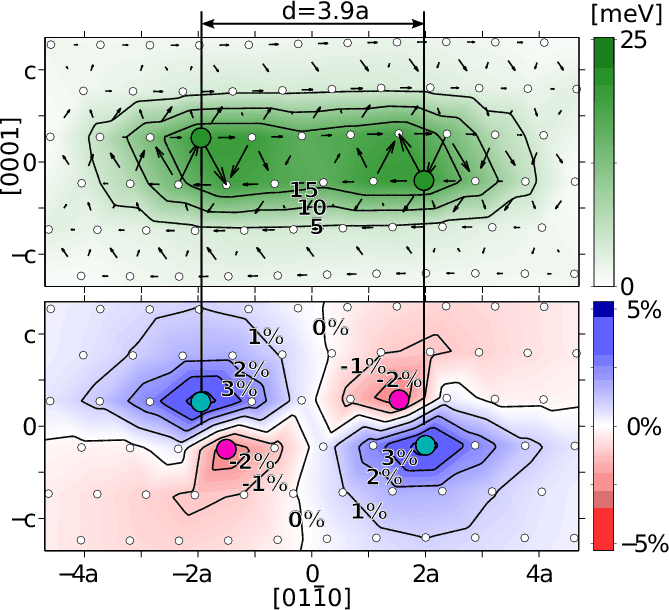}

\caption{Mg basal screw dislocation core geometry with atomically resolved
slip energy (top) and volumetric strain (bottom).  The equilibrium geometry
is found using first-principles flexible boundary condition methods.  The
circles show positions of atomic rows repeated out of the page (the
$[2\bar1\bar10]$ direction), while the magnitude of arrows between
neighboring sites represent the relative displacement of the neighboring
rows out of the page.  The arrow magnitude is maximum for the
\textit{partial} slip.  The dislocation splits into two partials---whose
cores are centered on a closed circuit of arrows---with screw components
(displacements out of the page) of $0.5a$, and edge components
(displacements in the horizontal slip plane) of $0.289a$, separated by
$3.9a$.  The slip energy is computed from the atomic geometry based on the
relative displacements of neighbors, and has a maximum value of 20meV at
the green sites.  The volumetric strain is computed from the average change
in nearest neighbor distances, and has a maximum tensile value of $+4.6\%$
at the cyan sites and maximum compressive value of $-2.7\%$ at the magenta
sites.}
\label{fig:screw}
\efig

\bfig
\centering
\includegraphics[width=\smallfigwidth]{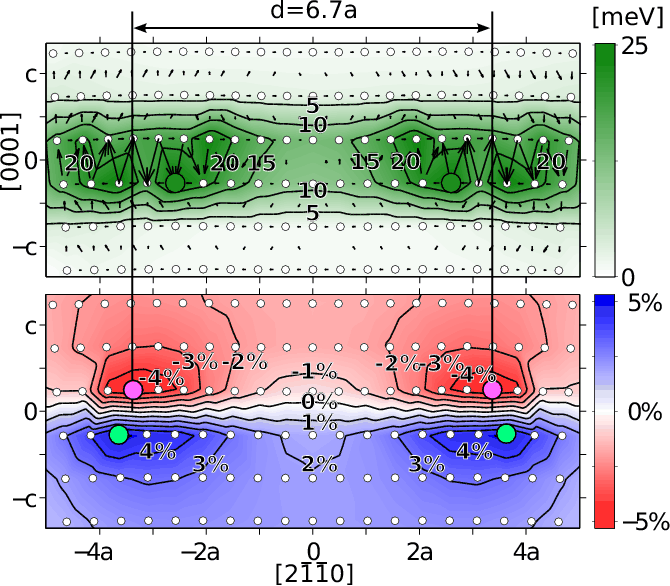}
\caption{Mg basal edge dislocation core geometry with atomically resolved
slip energy (top) and volumetric strain (bottom).  The equilibrium geometry
is found using first-principles flexible boundary condition methods.  The
circles show positions of atomic rows repeated out of the page (the
$[01\bar10]$ direction), while the magnitude of arrows between neighboring
sites represent the relative displacement of the neighboring rows along the
horizontal slip plane.  The arrow magnitude is maximum for the
\textit{partial} slip.  The dislocation splits into two partials---whose
cores are centered on a closed circuit of arrows---with edge components
(displacements in the horizontal slip plane) of $0.5a$, and screw
components (displacements out of the page) of $0.289a$, separated by
$6.7a$.  The slip energy is computed from the atomic geometry based on the
relative displacements of neighbors, and has a maximum value of 25meV at
the green sites.  The volumetric strain is computed from the average change
in nearest neighbor distances, and has a maximum tensile value of $+5.3\%$
at the cyan sites and maximum compressive value of $-4.9\%$ at the magenta
sites.}
\label{fig:edge}
\efig

\fig{screw} and \fig{edge} show the first-principles equilibrium basal
screw and edge dislocation core geometries analyzed in terms of size and
slip at each atomic site.  The screw dislocation has displacements (slip)
parallel to the line direction (out of the page), and the edge dislocation
has slip perpendicular to the line direction, representing the two limiting
cases for basal dislocation geometry.  The equilibrium screw dislocation
geometry dissociates into $\frac{a}{3}[10\bar10]$ and
$\frac{a}{3}[01\bar10]$ partial dislocations separated by $3.9a$ of I2
stacking fault, where $a$ is the basal lattice spacing.  The equilibrium
edge dislocation also dissociates into partial dislocations (half the slip
of the full dislocation), separated by $6.7a$ of stacking fault.  The
``cores'' of the partials have the largest change in local geometry from
bond-stretching and bond-bending.  For a screw dislocation, there is no
far-field dilation (tension or compression); however, we find a large
volumetric strain in the partial cores.  The strain is comparable to that
in the edge dislocation core, which \textit{does} have a far-field dilation
strain.  We quantify the bond-bending with the relative displacement $\vec
d$ of each atomic row, and average the energy of stacking faults with
displacement $\vec d$ of the six neighbors to give a slip energy
$E_\text{slip}$.  This energy is localized to the partial cores and the
stacking fault separating them.  A solute's binding energy will change in
the dislocations due to the different local geometry; moreover, as the
cores show the greatest change in local environment, we expect solutes to
provide the strongest interactions from sites within the dislocation cores.

\bfig
\centering
\includegraphics[width=\smallfigwidth]{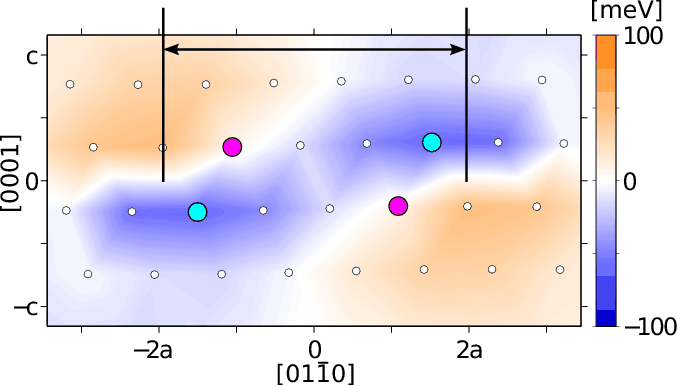}\\
\includegraphics[width=\smallfigwidth]{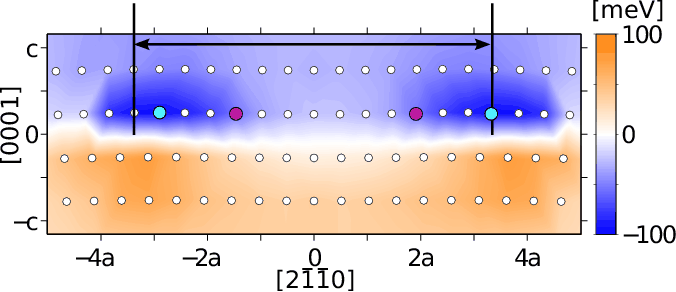}

\caption{Al interaction energies with Mg screw (top) and edge (bottom)
dislocations, from first principles.  We substitute Al atoms at different
sites in and around the partial cores, and compute the energy differences
after relaxation.  The cyan sites show points of maximum binding energy
while the magenta sites show maximum interaction force---i.e., greatest
change in binding energy along the slip plane (c.f.~\tab{Al} for numerical
values).  The sites between the partial cores and stacking faults have the
largest interaction force for both dislocation types.  The strong Al
interaction with the screw dislocation is surprising, and can only be
accurately resolved through the atomic-scale calculation of the full
dislocation core geometry.}
\label{fig:direct}
\efig

\fig{direct} shows the first-principles interaction energy---the change in the
solute binding energy---for an Al solute with screw and edge dislocation
cores.  We find the strongest interaction at the site of maximum
compression near the center of the dislocation partial cores for both the
edge and the screw geometries.  We expect this, since Al is smaller than Mg
and is affected by the size changes in the core.  The interaction force
(the derivative of the interaction energy in the slip direction) is largest
between the partial cores and the stacking fault for both dislocations,
with nearly equal magnitudes for edge and screw.  The similar values for
the screw and edge forces are surprising as elasticity theory predicts a
very weak far-field interaction for the screw
dislocation\cite{HullBacon2001}.  If we apply anisotropic elasticity theory
at a distance of $c/4$ from the slip plane (where $c/2$ is the distance
between parallel basal slip planes) to the screw and edge dislocations
separated into mixed partials ($\frac{a}{3}[1\bar100]$ and
$\frac{a}{3}[10\bar10]$), anisotropic elasticity theory predicts maximum
and minimum volumetric strains in the partial cores of $\pm 7.2\%$ (screw)
and $\pm 12.7\%$ (edge).  This is significantly different from our
atomic-scale values of --2.7\%, +4.6\% (screw) and --4.9\%, +5.3\% (edge).
Moreover, elasticity theory predicts the maximum interaction forces for
solutes to be the maximum interaction energy divided by 2\AA, and that the
solute/screw interaction force will be $58\%$ of the solute/edge
interaction force.  All of the elasticity predictions contradict our
atomic-scale calculations: equal solute/screw and solute/edge interaction
forces, energy to force ratios of 6--8\AA, and smaller and unbalanced
maximum and minimum volumetric strains.  All of these differences highlight
the sensitivity of the solute-dislocation interaction to the partial core
geometries and the need for a first-principles approach to compute
stress-free atomic-scale dislocation core geometries.

We replace the elasticity approach to solute-dislocation interaction with a
fully atomic-scale approach based on size and chemical misfits.  The
long-range strain and stress fields of a dislocation are known from
anisotropic elasticity\cite{Bacon1979}, and so the change in local volume
around a solute (size misfit $\size$, the logarithmic derivative of the
lattice constant with solute concentration) controls the solute-dislocation
interaction.  In the dislocation core, the interaction energy is largest,
but it is no longer described by elasticity.  Despite this, we can use the
size misfit \textit{even in} the partial cores by computing the change in
local volume for each atomic site.  This approximation is the largest
contribution to the interaction energy; the next largest contribution is
from the slip in the partial cores and the stacking fault.  In the same way
that we determine a slip energy in the core of the dislocation, we find
that solutes change the response of the crystal to slip (chemical misfit
$\chem$, the logarithmic derivative of the stacking-fault energy with
solute concentration).  The chemical misfit determines how the
atomically-resolved \textit{slip energy} (c.f.,
\fig{screw} and \fig{edge}) will change with the addition of a solute.
Adding this contribution to the size misfit produces an accurate
approximation of the solute-dislocation interaction energy using the
dislocation core geometry.

\begin{table}[ht]
\caption{Al interaction energy with Mg basal screw and edge dislocations from
direct solute substitution calculations (c.f. \fig{direct}) and misfit
approximations based on the dislocation geometry.  The maximum binding
energy (meV) and maximum interaction force (meV/\AA) determine the
attraction of solutes to dislocations and solid-solution strengthening.
The misfit approximation uses two misfits with the atomic-scale
dislocation geometry: change in local volume, and bond-bending from slip to
compute interaction energies for solute with the dislocation core.  For
both dislocation geometries, the misfit approximation correctly captures
the interaction energies and forces compared with the more computationally
intensive direct calculation.}
\label{table:Al}
\begin{center}
\begin{tabular}{l@{\quad}c@{\quad}c}
\textbf{screw}
&$E_\text{binding}$
&$F_\text{max}$\\
\hline
direct (\fig{direct})		&60	&11.4\\
misfit: volume			&46	&12.6\\
misfit: volume + slip		&65	&11.2\\
\textbf{edge}
&$E_\text{binding}$
&$F_\text{max}$\\
\hline
direct (\fig{direct})		&99	&12.2\\
misfit: volume			&81	&11.6\\
misfit: volume + slip		&105	&11.5
\end{tabular}
\end{center}
\end{table}

\tab{Al} compares the solute-dislocation interaction for Al computed using
direct substitution into the dislocation cores with our combined size and
chemical misfit approach.  The response of Al to changes in local volume
and slip capture most of the interaction energy and forces in the
dislocation core.  Moreover, the strain and slip energies are taken
directly from the equilibrium core geometries, and the size and
chemical misfit calculations include the local response of Mg atoms
neighboring the Al solute.  The size-misfit approximation is expected to be
accurate in the far-field, and is accurate even in the partial cores where
the interaction is the strongest.  The slip energy is needed to represent
the stacking fault region between the partials, and the center of the
partial cores themselves.  The maximum interaction forces from the misfit
calculation are accurate to within 5\%\ of the direct calculations for both
screw and edge geometries, with deviations of only 5meV for the interaction
energies; hence, we can predict the interaction energies of other solutes
by using the size and chemical misfits combined with the equilibrium
dislocation core geometries.  This permits us to use much simpler and
computationally efficient first-principles calculations of misfits with our
first-principles calculation of pure Mg dislocation cores.

\bfig
\centering
\includegraphics[width=\smallfigwidth]{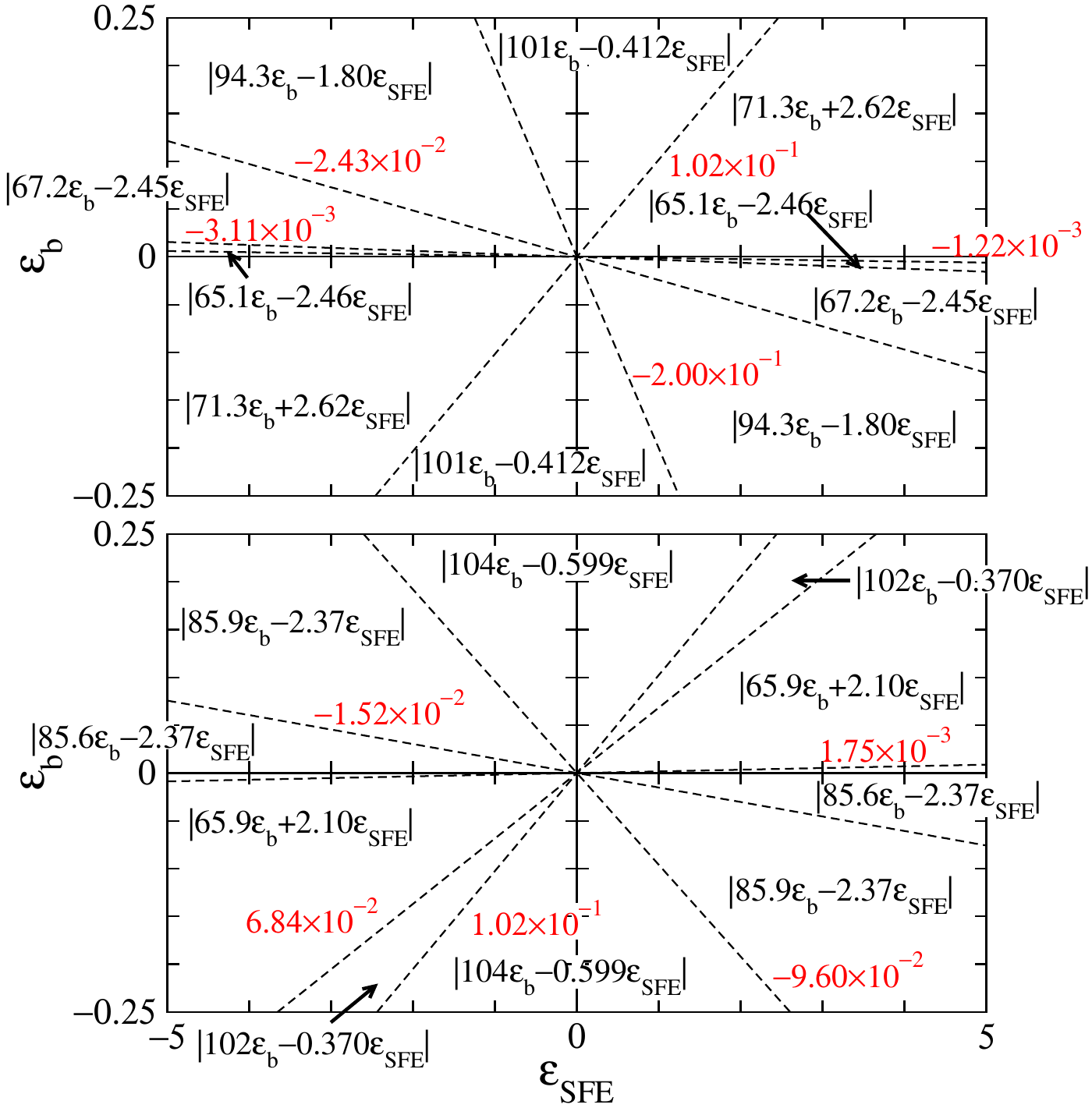}

\caption{Formulae for computation of maximum solute-dislocation interaction
forces from $\size$ and $\chem$ for edge (top) and screw (bottom)
dislocations.  Each solute substitutional site in the screw or edge
dislocation has a local volumetric strain and slip energy; we use the size
and chemical misfits to approximate the solute-dislocation interaction
energy by linearly scaling those energies.  The derivative of the
solute-dislocation interaction energy along the slip direction gives the
interaction force, and the maximum interaction force appears as input in
our strengthening model (\eqn{model}).  The particular site that produces
the largest interaction force changes depending on the magnitudes of the
misfits $\size$ and $\chem$, leading to different regimes.  The slopes of
the lines separating the different regimes are indicated in red.}
\label{fig:misfits}
\efig

\fig{misfits} provides the formulae for the maximum solute-dislocation
interaction force in terms of misfits.  The interaction force from
the first-principles solute binding energy is $F_\text{max} = \max|\hat m
\cdot\nabla E_\text{binding}|$ where $\hat m$ is the dislocation slip
direction ($[1\bar100]$ for screw and $[11\bar20]$ for edge), and
$E_\text{binding}$ is the sum of the solute slip and size interaction
energies from \eqn{misfitbinding}.  The gradients of the solute size and
slip interactions are calculated as a centered difference in the slip
direction for sites near the partial dislocation cores where the gradient
is the largest.  The gradients were scaled by the size and chemical misfits
for each solute and summed to determine the maximum interaction force.  The
end result is that, for any combination of $\size$ and $\chem$, the
strongest interaction force site is known, and its pinning force can be
computed as a linear combination of the two parameter; however, the
coefficients change as different magnitudes of $\size$ and $\chem$ will
select different sites as the strongest interaction force site.
\fig{misfits} graphically summarizes all of the formulae for each range of
$\size$ and $\chem$.  For example, Al has $\size=-0.115$ and
$\chem=-1.252$; then, for the edge dislocation, the interaction force is
given by the formula in the southwest corner $|71.3\size+2.62\chem| =
11.5\text{meV/\AA}$, and for the screw dislocation, the interaction force
is given by the formula in the narrow southwest wedge
$|102\size-0.370\chem|=11.2\text{meV/\AA}$.  For Zn, $\size=-0.153$ and
$\chem=+0.317$, so the edge dislocation interaction force is given by the
formula in the south section $|101\size-0.412\chem|=15.6\text{meV/\AA}$ and
for the screw dislocation, the interaction force is given by the formula in
the south section $|104\size-0.599\chem| = 16.1\text{meV/\AA}$.

%%%%%%%%%%%%%%%%%%%%%%%%%%%%%%%%%%%%%%%%%%%%%%%%%%%%%%%%%%%%%%%%%%%%%%%%
\TITLE{Solid-solution strengthening model}

To predict solid-solution strengthening, we use our first-principles
atomic-scale solute-dislocation interaction calculation as input to a
dilute-concentration, weak-obstacle model for solid-solution strengthening
from Fleischer\cite{fleischer1964}.  As the dislocation moves in the slip
plane under stress, it encounters randomly placed immobile solute atoms
each of which provides a ``pinning'' force up to the maximum
solute-dislocation interaction force $F_\text{max}$.  This point-pinning
model is applicable for isolated obstacles with a short-ranged---on the
scale of the solute-separation distance---interaction force between solute
and dislocations, and hence will capture the dilute-concentration limit.
It also assumes a random solute distribution that does not rearrange due to
the dislocation strain-fields; hence, it is operable at temperatures where
no appreciable solute diffusion occurs.  The nearly straight dislocation
bows at the solute until it reaches a critical bowing angle $\theta_c$ and
the line tension $\scrE$ pulls the dislocation past the solute.  At the
critical angle $\theta_c$, $F_\text{max} = 2\scrE \sin((\pi -
\theta_c)/2)$.  For weak obstacles, the dislocation is nearly straight, and
the critical bowing angle is only slightly smaller than $\pi$, hence $\pi -
\theta_c \approx F_\text{max}/\scrE \ll 1$.  The mean distance between
pinning points, $L$, is the average distance between nearest randomly
placed solutes in a circular wedge of angle $\alpha = (\pi -
\theta_c)/2$, where $L = \sqrt{\pi/(2\rho_s\alpha)}$, and $\rho_s$ is the
density of solutes in the slip plane.  The solute density is $2\conc$
(atomic concentration $\conc$) per $\sqrt{3} b^2/2$ (Burgers vector $b=a$,
basal lattice constant), assuming that the solute can appear either above
or below the slip plane; hence,
\beu
L = \left(\frac{\pi}{\rho_s (\pi - \theta_c)}\right)^{1/2}
= \left(\frac{\pi\sqrt{3} b^2}{4(\pi - \theta_c)\conc}\right)^{1/2}
= \frac{(3\pi^2)^{1/4}b}{2}\left(\frac{\scrE}{F_\text{max}}\right)^{1/2}
\conc^{-1/2}.
\eeu
That is, the distance between pinning points increases with decreasing
solute concentration $\conc$ and with decreasing interaction force.
Given the mean solute spacing, the strengthening (increase in
critical-resolved shear stress in the basal plane necessary to overcome the
solute restraining force) is:
\beu
\dtCRSS = 
\frac{F_\text{max}}{bL} =
\frac{2}{(3\pi^2)^{1/4}}
\frac{\scrE}{b^2}
\left(\frac{F_\text{max}}{\scrE}\right)^{3/2}\sqrt{\conc},
\eeu
which scales as $F_\text{max}^{3/2}$ and $\sqrt{\conc}$.  This is the
change for a (nearly) straight dislocation line of a given character
ranging from edge to screw; the interaction energy and line tension
correspond to the particular line orientation.  The strengthening of an
average dislocation loop---which continuously ranges from edge to screw
orientations---is weighted by the line tension.  The shape of a basal loop
can be estimated as an ellipse with axial ratio $\alpha = K_e/K_s$, where
$K_e=25.6\text{GPa}$ and $K_s=18.6\text{GPa}$ are the edge and screw line
energy prefactors\cite{HirthLothe} as determined from the first-principles
elastic constants $C_{ij}$.  For Mg, this gives a weight of 0.539 for the
screw dislocation and 0.461 for the edge.  The screw and edge line tensions
are $\scrE_s = \frac{1}{2}K_sb^2 = 591\text{meV/\AA}$ and $\scrE_e =
\frac{1}{2}K_eb^2 = 817\text{meV/\AA}$; so, the dislocation loop
weighted-average strengthening for each solute is
\be
\dtCRSS = \left[
0.30 \Big(F^\text{screw}_\text{max}\Big)^{3/2} + 
0.22 \Big(F^\text{edge}_\text{max}\Big)^{3/2}
\right]\cdot \conc^{1/2},
\label{eqn:model}
\ee
where $\dtCRSS$ is in MPa and the interaction forces are in meV/\AA.

Hence we need to consider the contributions to strength from \textit{both}
edge and screw dislocations for all solutes in our modified-Fleischer
model.  The concentration-independent prefactor in \eqn{model} is the
solute ``strengthening potency.''  The solubility of a solute controls the
maximum possible $\sqrt{\conc}$, and hence the maximum possible
strengthening of a solute in an alloy.  Combining the calculation of
interaction strengths from their misfits, the strength potency can be fit
to an approximate functional form
\beu
\dtCRSS \approx (38.9\text{MPa})\left\{
\left(\frac{\size}{0.176}\right)^2
+\left(\frac{\chem}{5.67}\right)^2
-\frac{\size\chem}{2.98}
\right\}^{3/2}\cdot \conc^{1/2}.
\eeu
The negative $\size\chem$ term shows that the effect of the size and
chemical misfits is not completely independent, but that due to the
geometry of the dislocation core---where the sites of largest slip energy
are neighboring sites of largest strain energy---a larger gradient can be
produced by having the size and slip misfits with opposite signs.  However,
the magnitude of the denominators also shows that the size misfit is the
dominant interaction.  The fit had a maximum error of 10\%, with the
average error being 3\%.  As this model ignores thermally activated
processes to overcome solute obstacles, it is appropriate at temperatures
without appreciable diffusion, but the fundamental interaction data is
usable for high temperature strengthening models as well.  All of the
material information that enters \eqn{model} comes \textit{completely} from
first principles: crystal structure, lattice and elastic constants,
dislocation core geometries, and solute misfits.

\bfig
\centering
\includegraphics[width=\smallfigwidth]{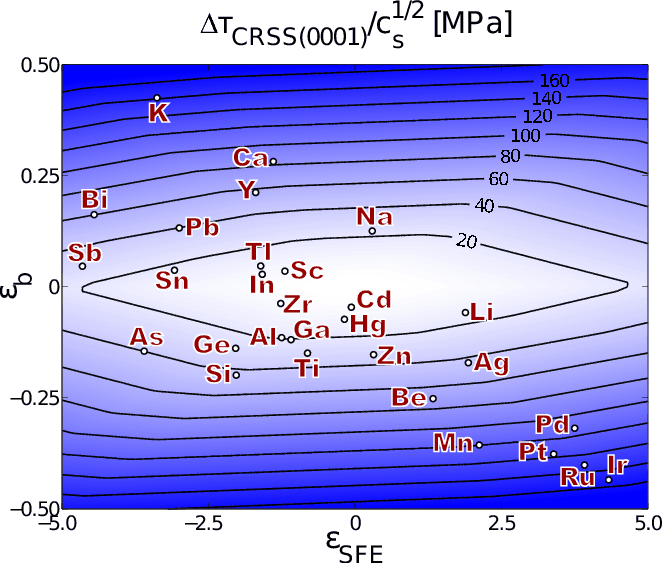}

\caption{Solid-solution strengthening potency (contours) versus their size
($\size$) and chemical ($\chem$) misfits for 29 different solutes in Mg.
The misfits, combined with the screw and edge dislocation geometries
(c.f. \fig{screw} and \fig{edge}) determine the maximum interaction forces.
In the dilute limit for weak obstacles, the strengthening---characterized
by a change in the basal critical-resolved shear stress $\dtCRSS$---scales
with $\sqrt{\conc}$ and with the interaction forces to the 3/2 power in
\eqn{model}.  We can efficiently predict the solute strengthening for a
whole range of solutes from the first-principles core geometry.  This
``design map'' shows solutes with comparable strengthening potency, and
also gives a guide to solubility---which decreases with increasing misfit
magnitudes.}
\label{fig:design}
\efig

\fig{design} combines the first-principles data for size and chemical misfits
with the simple solid-solution strengthening model of \eqn{model} into a
``design map'' for the strengthening potencies of 29 different solutes.
The contours show equal strengthening potency versus size and chemical
misfits.  The size and chemical misfits can be easily computed for any
substitutional solute in the periodic table, and then the screw and
edge dislocation maximum interaction forces to predict the change in
critical-resolved shear stress for basal slip from \eqn{model}.  This map
provides a rational method to select equipotent solutes to replace less
``desirable'' elements based on high mass, cost, or other processing
concerns.  In addition, the misfits also give rough qualitative information
about solubility, as larger misfit magnitudes lead to lower solubilities.
For example, for Zn with a potency $\dtCRSS/\sqrt{\conc} = 32.5\text{MPa}$,
the corresponding small magnitudes of $\size$ and $\chem$ suggest a high
solubility.  Alternatively, the much larger potencies of Ir (172MPa) and K
(161MPa) are due to the larger magnitudes of $\size$ and $\chem$, giving
rise to a much lower solubility.  Yttrium represents a reasonable
compromise between strength and solubility.

\newcommand*{\entry}[1]{\multicolumn{1}{c}{\textbf{#1}}}
\newcommand*{\dentry}[1]{\multicolumn{2}{c}{\textbf{#1}}}
\newcommand*{\tabcite}[1]{~\protect\cite{#1}}
\begin{table*}[ht]
\caption{Substitutional solutes with pseudopotential valence configuration
and energy cutoff, size and chemical misfits, corresponding maximum
interaction force with an edge and screw dislocation, and computed
interaction forces from \fig{misfits} and strengthening potency from
\eqn{model}, and from single crystal, low temperature, dilute-concentration
experimental measurements where available.  In the case of Bi and Pb, low
concentration data is not available; instead plateau stress (potency at
higher concentration) serves as a lower limit on the dilute-concentration
potency.}
\label{table:solute}
\begin{center}
\renewcommand{\arraystretch}{0.6} %% instead of \squeezetable
\renewcommand{\tabcolsep}{6pt}
\begin{tabular}{@{}clc|rr|cc|cc@{}}
\entry{}&
\entry{}&
\entry{}&
\entry{}&
\entry{}&
\dentry{F$_\text{max}$ (meV/\AA)}&
\dentry{potency (MPa)}\\
\entry{solute}&
\dentry{\quad USPP \hfill cutoff (eV)}&
\entry{$\size$}&
\entry{$\chem$}&
\entry{edge}&
\entry{screw}&
\entry{eqn.\ref{eqn:model}}&
\entry{exper.}\\[1pt]
\hline
Ag	&[Kr]$4d^{10}5s^1$                &235	&--17.1\%	& 1.93  	&19.6	&19.3& 44.1&\\
Al	&[Ne]$3s^23p^1$                   &168	&--11.5\%	&--1.25  	&11.5	&11.2& 19.6&21.2\tabcite{Akhtar1972}\\
As	&([Ar]$3d^{10}$)$4s^24p^3$        &188	&--14.5\%	&--3.60  	&19.8	&17.1& 40.3&\\
\hline
Be	&[He]$2s^2$                       &327 	&--25.2\%	& 1.33  	&26.1	&27.1& 71.1&\\
Bi	&([Xe]$4f^{14}5d^{10}$)$6s^26p^3$ &138	& 16.2\%	&--4.45  	&23.3	&24.5& 60.6&$>25.0$\tabcite{Akhtar1972}\\
Ca	&[Ar]$4s^2$                       &138 	& 28.2\%	&--1.39  	&29.1	&30.3& 83.8&\\
\hline
Cd	&[Kr]$4d^{10}5s^2$                &218	& --4.6\%	&--0.06  	& 4.6	& 4.8&  5.3& 6.0\tabcite{Scharf1968}\\
Ga	&([Ar]$3d^{10}$)$4s^24p^1$        &169	&--11.9\%	&--1.09  	&11.6	&11.8& 20.7&\\
Ge	&([Ar]$3d^{10}$)$4s^24p^2$        &181	&--13.9\%	&--2.04 	&15.2	&13.4& 27.6&\\
\hline
Hg	&([Xe]$4f^{14}$)$5d^{10}6s^2$     &207	& --7.3\%	&--0.18  	& 7.3	& 7.5& 10.5&\\
In	&([Kr]$4d^{10}$)$5s^25p^1$        &138	&  2.8\%	&--1.58  	& 5.8	& 6.2&  7.6& 9.0\tabcite{Akhtar1972}\\
Ir	&([Xe]$4f^{14}$)$5d^76s^1$        &258	&--43.5\%	& 4.33  	&48.8	&48.0&173.1&\\
\hline
K 	&[Ar]$4s^1$                       &138 	& 42.5\%	&--3.38  	&46.2	&46.4&162.5&\\
Li	&[He]$2s^1$                       &138 	& --5.8\%	& 1.89  	& 8.9	& 9.4& 14.4&11.2\tabcite{Yoshinaga1963}\\
Mn	&[Ar]$3d^64s^1$                   &295	&--35.6\%	& 2.12  	&37.4	&38.5&120.8&\\
\hline
Na	&[Ne]$3s^1$                       &138 	& 12.6\%	& 0.29  	&12.6	&12.9& 23.6&\\
Pb	&([Xe]$4f^{14}5d^{10}$)$6s^26p^2$ &138	& 13.2\%	&--3.00  	&17.9	&18.5& 40.1&$>14.0$\tabcite{Akhtar1972}\\
Pd	&[Kr]$4d^95s^1$                   &259	&--31.8\%	& 3.75  	&36.7	&36.2&113.5&\\
\hline
Pt	&([Xe]$4f^{14}$)$5d^96s^1$        &249	&--37.7\%	& 3.39  	&41.6	&41.3&137.5&\\
Ru	&[Kr]$4d^75s^1$                   &265	&--40.1\%	& 3.92  	&44.9	&44.2&153.0&\\
Sb	&([Kr]$4d^{10}$)$5s^25p^3$        &139	&  4.6\%	&--4.65  	&14.5	&15.0& 29.3&\\
\hline
Sc	&[Ar]$3d^24s^1$                   &195	&  3.5\%	&--1.20  	& 5.5	& 5.9&  7.0&\\
Si	&[Ne]$3s^23p^2$                   &196	&--19.9\%	&--2.03  	&19.5	&19.6& 44.5&\\
Sn	&([Kr]$4d^{10}$)$5s^25p^2$        &138	&  3.7\%	&--3.08  	&10.1	&10.5& 17.1&24.3\tabcite{VanderPlanken1969}\\
\hline
Ti	&[Ar]$3d^34s^1$                   &236	&--14.9\%	&--0.81  	&14.8	&15.1& 29.8&\\
Tl	&([Xe]$4f^{14}$)$5d^{10}6s^26p^1$ &231	&  4.7\%	&--1.61  	& 7.3	& 7.8& 10.8& 8.2\tabcite{Levine1959}\\
Y 	&[Kr]$4d^25s^1$                   &155	& 21.2\%	&--1.70  	&23.1	&23.2& 57.3&\\
\hline
Zn	&[Ar]$4s^23d^{10}$                &272	&--15.3\%	& 0.32  	&15.6	&16.1& 32.7&31\tabcite{Akhtar1969}\\
Zr	&[Kr]$4d^35s^1$                   &195	& --3.8\%	&--1.27  	& 6.0	& 5.1&  6.7&
\end{tabular}
\end{center}
\end{table*}

\tab{solute} provides misfits and potencies for all 29 solutes, as well as
comparisons to available experimental data (for Al, Zn, Bi, Cd, In, Li, Pb,
Sn, and Tl).  For comparison, we look to single-crystal, low temperature
measurements of critical resolved shear stress of Mg in the low
concentration limit.  For example, the available experimental data for
potencies of common solutes Al and Zn in Mg extrapolated to
0K\cite{Akhtar1972,Akhtar1969} give 21.2MPa and 31MPa, versus our
first-principles prediction of 19.5MPa and 32.5MPa---validating our
computation and modeling approach.  For Bi and Pb, only plateau stress data
is available---this is the strengthening divided by $\sqrt{\conc}$ at
larger concentrations than considered here.  We expect the plateau
strengthening coefficient to be a lower limit on our dilute concentration
potency, and so our data remains consistent with the available experimental
data.

%%%%%%%%%%%%%%%%%%%%%%%%%%%%%%%%%%%%%%%%%%%%%%%%%%%%%%%%%%%%%%%%%%%%%%%%
\TITLE{Conclusions}

The basal-slip strengthening design map for Mg represents an important
development in predicting chemical effects on strengthening in an accurate
and efficient manner.  In addition to predicting strength, we find a strong
solute interaction with screw dislocations equal to that of edge
dislocations: this demonstrates the need for first-principles
quantum-mechanical calculations with flexible boundary conditions to reveal
defect interactions.  Maximizing strength requires careful consideration of
the important tradeoff between high strengthening capacity (i.e., large
solute misfits) and high solubility (low misfit magnitudes).  Predicting
strength above dilute concentrations, at higher temperatures and including
other mechanisms such as forest strengthening\cite{Soare2008} will require
the fundamental input from the present solute strengthening predictions.
Moreover, cross-slip in Mg depends on the constriction of basal screw
dislocations, and should be aided by solutes with positive $\chem$ to
increase the basal stacking-fault energy.  By finding solutes that can
improve cross-slip while simultaneously strengthening basal slip---as
approached here---the strength anisotropy of Mg alloys can be reduced,
improving ductility.  The natural extension of the computational
methodology developed herein is to predict solute effects on slip in
prismatic planes, including increasing cross-slip for ductility
enhancement.  In addition, atomically-resolved strengthening predictions
hold great promise for the design of other technologically important HCP
metals such as Ti and Zr.

\subsection*{Acknowledgments}
The authors thanks W.~A.~Curtin for helpful discussions.  This research was
sponsored by NSF through the GOALI program, Grant 0825961, and with the
support of General Motors, LLC.  This research was supported in part by the
National Science Foundation through TeraGrid resources provided by NCSA and
TACC, and with a donation from Intel.  Computational resources, networking,
and support at GM were provided by GM Information Systems and Services.
\fig{geom} was rendered with VMD\protect\cite{Humphrey1996}.

% \bibliographystyle{model3-num-names}
% \bibliography{references}

\end{document}